\documentstyle[epsfig,psfig]{aipproc}

\begin{document}
\title{A jet-disk symbiosis model for Gamma Ray Bursts: fluence 
distribution, CRs and $\nu$'s}

\author{G. Pugliese$^1$, H. Falcke$^1$, Y. Wang$^{2,3}$, 
P. L. Biermann$^1$}
\address{$^1$Max-Planck-Institut f\"ur Radioastronomie, Auf dem 
H\"ugel 69, 53121 Bonn, Germany\\
$^2$Purple Mountain Observatory, Academica Sinica, Nanjing 210008, China\\ 
$^3$National Astronomical Observatories, Chinese Academy of Sciences}

\maketitle

\vspace{0.8cm}

\paragraph*{(To appear in Proceedings of the 10th Annual October 
Astrophysics Conference in Maryland: Cosmic Explosions !)}

\begin{abstract}

We consider a jet-disk symbiosis model to explain Gam\-ma Ray Bursts
and their afterglows. It is proposed that GRBs are created inside a
pre-existing jet from a neutron star in a binary system which collapses 
to a black hole due to accretion. In our model we assume that a fraction 
of the initial energy due to this transition is deposited in the jet by 
magnetic fields. The observed emission is then due to an ultrarelativistic 
shock wave propagating along the jet. Good agreement with observational 
data can be obtained for systems such as the Galactic jet source SS433. 
Specifically, we are able to reproduce the typical observed afterglow 
emission flux, its spectrum as a function of time, and the fluence 
distribution of the corrected data for the 4B BATSE catalogue. We also 
studied the relation between the cosmological evolution of our model and 
the cosmic ray energy distribution. We used the Star Formation Rate (SFR) 
as a function of redshift to obtain the distribution in fluences of GRBs 
in our model. The fluence in the gamma ray band has been used to calculate 
the energy in cosmic rays both in our Galaxy and at extragalactic distances. 
This energy input has been compared with the Galactic and extragalactic 
spectrum of cosmic rays and neutrinos. We found that in the context of our 
model it is not possible to have any contribution from GRBs to either the 
extragalactic or the Galactic cosmic ray spectra. 

\end{abstract}

\section*{Introduction}

Gamma-Ray Bursts are short bursts that peak in the soft $\gamma$-ray band, 
between 100 KeV and a few MeV. The duration of their emission goes from $10 
\times 10^{-3}$ s to $10^3$ s, and they show variability of the order of 
$\rm{ms}$. They also show persistent emissions in the X, optical, infrared 
and radio bands (afterglow), a spatially isotropic distribution, and a 
nonthermal spectrum. It is believed that GRBs are associated with 
relativistic shocks caused by a relativistic fireball in a pre-existing gas, 
such as the interstellar medium or a stellar wind/jet, producing and 
accelerating electrons/po\-si\-trons to very high energies, which produce 
the gamma-emission and the various afterglows observed \cite{paczy86,rees93}. 
More than 30 years after their di\-sco\-ve\-ry, thanks to the Burst and 
Transient Source Experiment (BATSE) and the Italian-Dutch satellite 
BeppoSax, the scientific community knows that Gamma Ray Bursts (GRBs) 
are isotropically distributed in the sky and that at least some of them 
are at cosmological distances. But the present data available for redshift 
position and host galaxy localization are still too few to give us good 
statistics to study the evolution of GRBs and their redshift distribution. 
Because of this lack of information, it is still necessary to assume that 
GRBs follow the statistical distribution of some other well known objets 
to obtain the GRBs fluence or flux distribution itself \cite{feni95,cohe95}. 

\section*{GRB jet model: key points}

In our model \cite{pugl99}, GRBs develop in a pre-existing jet. We consider 
a binary system formed by a neutron star and an O/B/WR companion in which 
the energy of the GRB is due to the accretion-induced collapse of the
neutron star to a black hole. To fix the jet parameters we use the basic 
ideas of the jet-disk symbiosis model by Falcke $\&$ Biermann \cite{falc95}. 
In this model, accretion disk, jet, and compact object are considered as an 
entire system. Mass and energy conservation are applied and the total jet 
power $Q_{\rm jet}$ is found to be a substantial fraction of disk luminosity 
$L_{\rm disk}$. We assume that the collapse of a neutron star to a black hole 
in a binary system induces a highly anisotropic energy release along the 
existing jet: a violent twist and jerk of the magnetic field. It initiates a 
relativistic shock wave, with an initial bulk Lorentz factor of about $10^4$. 
Baryonic mass is known to be low in jets. The bulk Lorentz factor evolution 
derives from the sweep up of the jet material. Magnetic field and particle 
number density evolution are obtained from the jump conditions in the 
ultrarelativistic shock. We consider a power law electron energy distribution 
with a low energy cut-off. Pre-existing energetic electrons/positrons are 
further accelerated in the shock. The afterglow emission is due to 
synchrotron and Inverse Compton processes from the shock region. The fluence 
of the initial burst is determined by shock, dissipation, and 
$\gamma$-$\gamma$ optical depth effects. The emission region is optically 
thin very early on and always in the fast cooling regime. There are only two 
parameters for the explosion: the energy in bulk flow along the jet, 
$E_{51} \cdot 10^{51} {\rm erg}$, and the fraction $\delta$ of shock energy 
in relativistic particles. The parameters from the binary system jet are: 
the mass flow $\dot M \cdot 10^{-5} {M_{\rm {\odot}}/{\rm{yr}}}$, the 
speed of the unperturbed jet $0.3 \cdot v_{0.3}$, as well as the minimum 
electron Lorentz factor $100 \cdot \gamma_{{\rm m},2}$. With these parameters 
and a distance $D_{28.5} \cdot 10^{28.5} {\rm{cm}}$, a time $t_5 \cdot 
10^5 \rm{s}$, and a frequency $\nu_{14} \cdot 10^{14} \rm{Hz}$, 
we obtain the correct flux level of the afterglow: 

\begin{equation}
F_{\rm {\nu}}^{\rm{(ob)}}(t) \simeq 7.45 \times 10^{-28} \delta 
(E_{51}^{5/4} \dot M_{-5 \rm j}^{-1/4} v_{0.3}^{1/4}) \gamma_{{\rm m},2} 
D_{28.5}^{-2} t_5^{-5/4} \nu_{14}^{-1} \enskip \rm{erg \enskip 
cm^{-2} s^{-1} Hz^{-1}} 
\end{equation}

\section*{Contribution to cosmic ray and neutrino flux}

\begin{figure}[b!] 
\epsfig{file=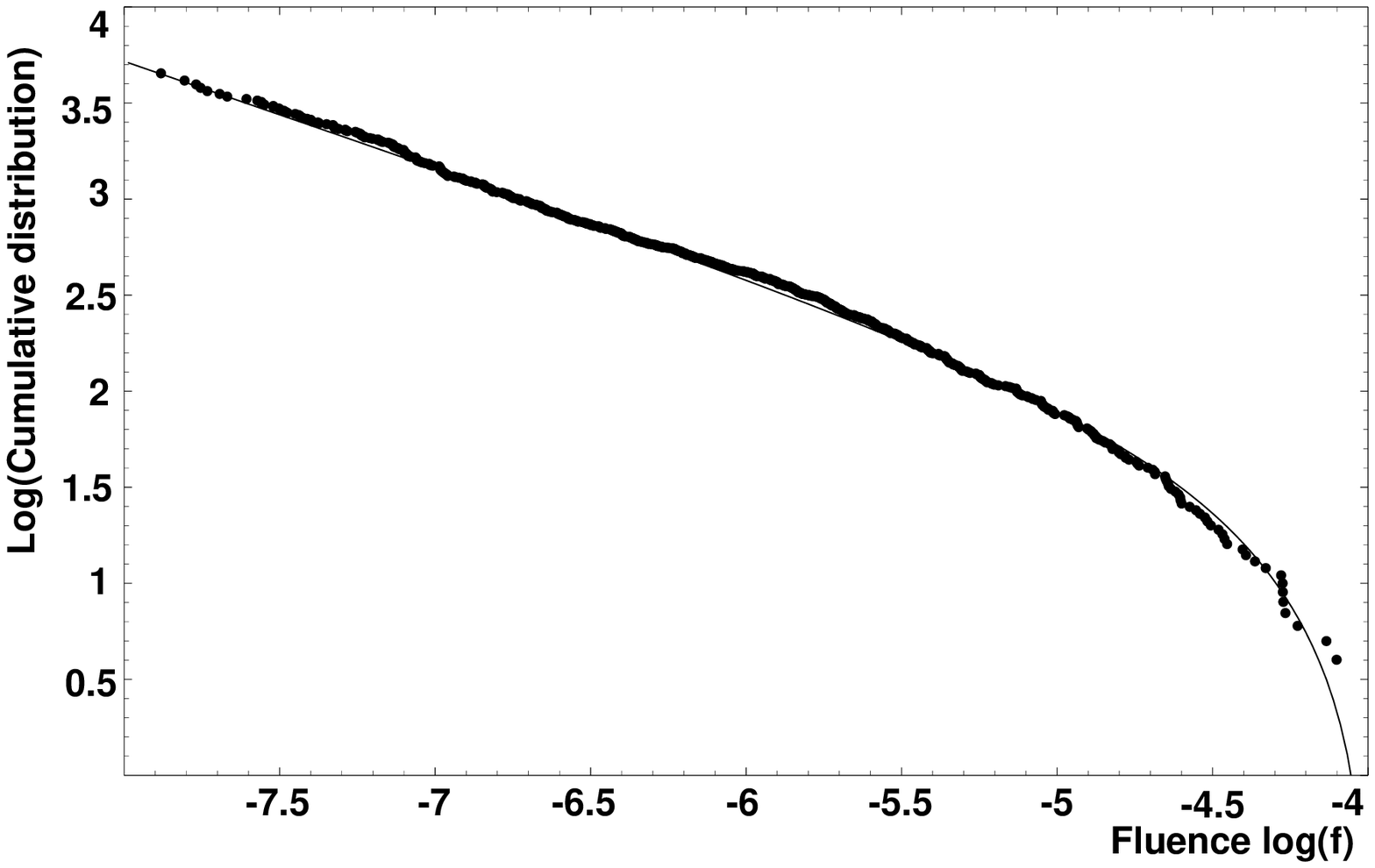,height=2.75in,width=2.75in}
\psfig{file=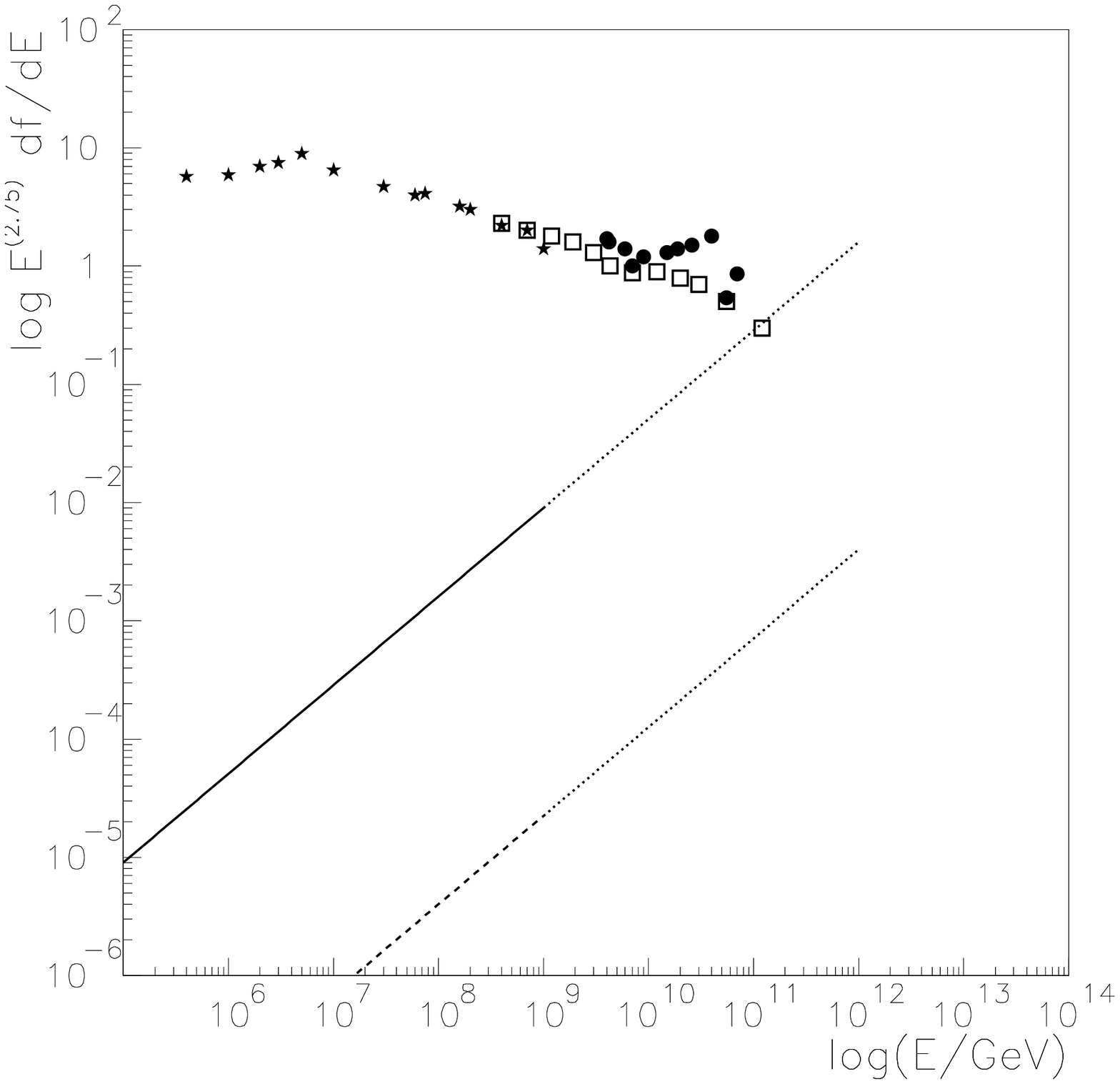, 
        bbllx=-0.01in, bblly=2.30in, bburx=3.4in, bbury=4.0in,scale=0.41}
\vspace{16pt}
\caption{\label{figura}Left: comparison between the cumulative fluence 
         distribution obtained with our model (solid line) and the corrected 
         data  of catalogue BATSE 4B (full circles), kindly provided by V. 
         Petrosian. The fit we obtained requires a power law luminosity 
         function distribution for GRBs $\phi(f) \, df = f^{-1.55} \, df$. 
         Right: comparison between the extragalactic GRB contribution and 
         the all particle cosmic ray spectrum, expressed in $\rm {[GeV^{-1} 
         \; cm^{-2} \; s^{-1} \; sr^{-1}]}$. The solid line corresponds 
         to the assumption that each GRB gives the same contribution to 
         the CR spectrum; this model is excluded by observations. The 
         dashed line corresponds to the model in which each GRB gives a 
         contribution proportional to its own fluence. Any contribution 
         beyond $10^{18}$ eV is ruled out (dotted lines) considering the 
         attenuation due to the interaction with the microwave background. 
         The stars represent the cosmic ray data from the Akeno experiment, 
         the open squares are the Fly's Eye data and the full circles are 
         the AGASA data.} 
\label{fig2}
\end{figure}

We calculated the GRB rate and compared the corresponding cumulative 
distribution in fluence with the data. We used the SFR as a function of 
the redshift presented by Madau \cite{mada96} with a flatter SFR at high 
redshift to obtain the corresponding fluence distribution of GRBs with 
the redshift and to use their rate to study the eventual contribution 
of GRBs to the cosmic ray distribution, both in our Galaxy and in the 
extragalactic region. We checked if in our jet model GRBs were standard 
candles. The corrected data for the 4B BATSE catalogue fluence distribution 
\cite{petr99} require the adoption of a luminosity function with a power 
$\phi(f) \propto f^{-1.55}$. The result of our calculations is shown in 
Fig.~\ref{figura}(left), in which the theoretical fluence distribution 
curve is compared with the 4B corrected data. Considering the total 
number of GRBs in BATSE catalogue, an observing time of 8 years, a volume 
scale of $h^{-3} 10^{10.8} \rm {Mpc^3}$, with $H_0 = h \, (100 \; {\rm km 
\; s^{-1} \; Mpc^{-1}})$ the Hubble constant, a beaming factor ${{4 \pi} 
\over {2 \pi \theta^2}} = 200 \, \theta^{-2}_{-1 \rm j}$, with $\theta$ 
the jet opening angle, the rate of GRBs is: 

\begin{eqnarray}
10^{-5.4} (h^3 \theta_{-1 \rm{j}}^{-2}) \; {\rm {GRBs \; per \; year \; 
per \; 100 \; Mpc^3}} 
\end{eqnarray}

We used the GRB rate obtained with the SFR from Madau and two different 
approaches to calculate the contribution from GRBs to the cosmic 
rays and the neutrino spectra. First we considered that each GRB gives 
the same contribution equal to $10 \%$ of the initial energy, here $10^{51} 
\rm{ergs}$. Secondly we assume that each GRB contributes proportionally 
to its own fluence; the fluence distribution adopted has a power law. 
In Fig.~\ref{figura}(right) we compared the all particle energy spectrum 
as measured by different ground-based experiment with the spectrum from 
GRBs in the case that each of them gives the same contribution (dashed 
line) and with the one in which the contribution is proportional to the 
fluence (solid line) for the extragalactic case. In the jet-disk symbiosis 
model for GRBs any extragalactic origin at high energies for cosmic rays is 
ruled out considering that for energies greater than $10^{18}$ eV (dotted 
line), the interactions with the microwave background are relevant and 
decrease the curve substantially. 
A corresponding analysis for the cosmic ray contribution from GRBs inside 
our Galaxy leads to the same result: Near $10^{18}$ eV the arrival directions 
of CRs are observed to be ispotropic to an excellent approximation,
and yet their diffusion time out of the Galaxy is much shorter than the 
time scale between GRBs in our Galaxy. Therefore the time for isotropization 
is not available, ruling out any contribution from GRBs. 

\section*{Conclusions}

To summarize, our model can explain the initial gamma ray burst, the
spectrum and temporal behaviour of the afterglows, the low baryon load,
an optical rise, and do all this with a modest energy budget. Moreover,
this GRB model is developed within an existing framework for galactic jet
sources, using a set of observationally well determined parameters.
Using a relatively small set of parameters, the jet-disk symbiosis model 
applied to GRBs, a tested SFR, and the fundamental physics 
of the photohadronic interactions we arrive at the conclusion that GRBs 
are unlikely to give any contribution to the high energy cosmic ray 
spectrum both inside and outside our Galaxy and to the neutrino spectrum 
as well.

\end{document}